\documentclass[12pt, letterpaper]{article}
\usepackage[utf8]{inputenc}
\usepackage{graphicx} 
\usepackage{cite}
\usepackage{amsmath}
\usepackage{amsfonts}

\title{An affine Weyl group characterization of polynomial Heisenberg algebras}
\author{V.S. Morales-Salgado}

\begin{document}
\maketitle
\begin{abstract}
We study deformations of the harmonic oscillator algebra known as polynomial Heisenberg algebras (PHAs),
and establish a connection between them and extended affine Weyl groups of type $A^{(1)}_m$, 
where $m$ is the degree of the PHA.
To establish this connection, we employ supersymmetric quantum mechanics to first connect a polynomial Heisenberg algebra 
to symmetric systems of differential equations.
This connection has been previously used to relate quantum systems to non-linear differential equations;
most notably, the fourth and fifth Painlev\'e equations. 
Once this is done, we use previous studies on the B\"acklund transformations of Painlev\'e equations 
and generalizations of their symmetric forms characterized by extended affine Weyl groups.
This work contributes to better understand quantum systems and the algebraic structures characterizing them.
\end{abstract}

\section{Introduction}
Supersymmetric quantum mechanics relates the eigenvalue problem of multiparametric families of quantum systems \cite{w81,w82}.
This tool is particularly powerful in that it allows the use of algebraic methods to study quantum systems \cite{m84,n84}.
In this regard, a supersymmetric transformation produces a realization of the factorization method in quantum mechanics \cite{f31,d57,f10}. 

The most common use of supersymmetric quantum mechanics has been the so-called spectral design \cite{f10,cks95,cf08,ai12}.
It consists in obtaining quantum systems with prescribed energy spectrum, 
departing from a system whose energy spectrum is already known
\cite{ais93,aicd95,bs97,fgn98,fhm98,jr98,mnn98,qv99,s99,mnr00,crf01,ast01,fs03,mr04,cfnn04,ff05,fs05,m09,q11,bf11a,m12,ggm13}.
Hamiltonians thus related are known as \emph{supersymmetric partners} and 
the relations among them are known as \emph{supersymmetric transformations}, 
\emph{Darboux transformations}, or \emph{intertwining relations} \cite {s99}.

Indeed, the algebraic structures characterizing each supersymmetric partner are also related.
One can be regarded as a deformation of the other.
The harmonic oscillator and its supersymmetric partners are frequent examples.
While the former is characterized by the oscillator algebra,
the later are characterized by polynomial Heisenberg algebras (PHAs).

In general, polynomial Heisenberg algebras are deformations of the oscillator algebra,
where the commutator between the ladder operators is a polynomial of the hamiltonian.
The degree of this polynomial is known as the degree of the PHA.
With regards to the eigenvalue problem of the hamiltonioans of PHAs, 
it has been found that their energy spectra consists of a superposition of equidistant \emph{ladders}
\cite{cfnn04,fh99,fnn04,mn08}.

It is also well known that systems ruled by PHA's of second and third degree 
are connected to the fourth and fith Painlev\'e equations, respectively
\cite{mn08,sh92,vs93,ad94,dek94,ekk94,ain95,srk97,acin00}.
This connection has been used in two ways.
First, to produce concrete realizations of PHAs,
one can use an specific solution of such Painlev\'e equations.
However, Painlev\'e IV and V equations are second order nonlinear differential equations.
Thus, an alternative and simpler use of the aforementioned connection is in the opposite direction:
concrete realizations of second and third degree PHAs can be used to obtain solutions to the Painlev\'e IV and V equations, 
respectively \cite{cfnn04,bf11a,bf14}.

Significant work has already been done in generalizing and exploiting 
the groups of rational transformations of Painlev\'e equations 
and their connection with symmetric systems or dressing chains.
For example, 
in \cite{o86} the theory of symmetry transformations of Painlev\'e IV equations is developed.
In \cite{vs93} authors introduce the dressing chains as higher order generalizations of the fourth and fifth Painlev\'e equations.
In \cite{ad94} the author defines periodically closed sequences of transformations equivalent to second to sixth Painlev\'e equations.
Symmetries of Painlev\'e equations and special functions have also been connected in 
\cite{fou00,wh03,c03a,c03b,c06,cj14,fc08,ggm21,v18}.
In \cite{n04} the symmetries of Painlev\'e equations characterized by affine Weyl groups are given and properties studied.
More recently, in \cite{ggm21} complete classification of rational solutions of the fourth Painlev\'e equation and 
its higher order generalizations is provided by means of the Weyl group $A_{2n}$ characterization.
Work has also been done in \cite{bf11a,mq16} with respect to to connections between Painlev\'e equations and quantum systems.

The main objective of this work is 
to generalize the connection between polynomial Heisenberg algebras and Painlev\'e equations. 
The generalization rests on the structure of the rational transformations of Painlev\'e equations, 
and their generalization to the case of the affine Weyl group of type $A^{(1)}_m$ \cite{n04}.
These transformations are instances of the so-called B\"acklund transformations and they allow one to 
obtain new solutions of Painlev\'e equations from a given known solution.
In short, the generalization of the structure of B\"acklund transformations of Painlev\'e equations 
to the case of the affine Weyl group of type $A^{(1)}_m$ can be used 
to study PHAs of arbitrary degree by means of tools provided by supersymmetric quantum mechanics. 

The following diagram illustrates the main idea of this work:
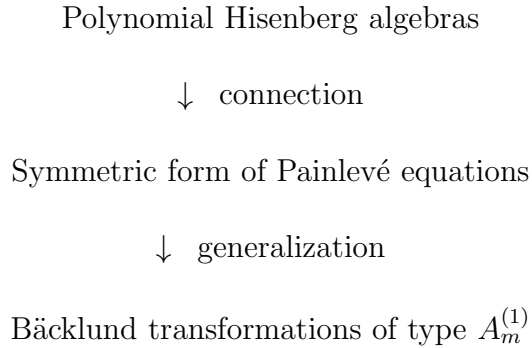
\begin{center}
 \begin{tabular}{c}
    Polynomial Hisenberg algebras \\ \\
    $\downarrow$ \, connection  \\ \\
    Symmetric form of Painlev\'e equations \\ \\
    $\downarrow$ \, generalization  \\ \\
     B\"acklund transformations of type $A^{(1)}_m$   
 \end{tabular}
\end{center}
The application of the first arrow is performed by supersymmetric quantum mechanics 
while the second arrow is performed by Nuomi's generalization of the structure of 
rational transformations of Painlev\'e equations \cite{ny98,ny99,n04}.

This article is organized as follows: 
In section 2 we briefly review the tool of supersymmetric quantum mechanics and its relation to the factorization method.
In section 3 we describe the object of this study, i.e., polynomial Heisenberg algebras.
In section 4 we present the manner in which low degree PHAs connect to symmetric forms of differential equations 
and generalize the results to the case of PHAs of arbitrary degree. 
Finally, in section 5 we give the final remarks on the results presented here.

\section{Supersymmetric transformations}
A \emph{supersymmetric transformation} in quantum mechanics relates two hamiltonians 
$H_i = -\frac{1}{2}\frac{{\rm d}^2}{{\rm d}x^2} +V_i(x)$, $i=0,1$, 
through the following operator equation
\begin{equation}\label{intertwining}
 H_1Q^+ = Q^+H_0 \,,
\end{equation}
where $Q^+$ is a $k$-th order differential operator known as the \emph{intertwining operator}.
Units such that $\hbar=m=1$ are assumed.

Equation (\ref{intertwining}) relates the energy spectra of the system $H_1$ with the one of $H_0$ \cite{cks95,f10,cf08,ai12}.
This spectral relations are obtained by noticing that 
if $\psi_n(x)$ solves the stationary Schr\"odinger equation for $H_0\psi_n(x)=E_n\psi_n(x)$, 
then $\phi_n\propto Q^+\psi_n(x)$ solves $H_1\phi_n(x)=E_n\phi_n(x)$. 
Thus, there might exist values in the energy spectrum of $H_0$ that are also in the energy spectrum of $H_1$.
One must notice that, in order to obtain an appropriate supersymmetric transformation,  
both eigenvalue problems, for $H_0$ and $H_1$, must possess the same domain of definition and boundary conditions.

As a simple example consider the harmonic oscillator $V_0=x^2/2$.
The nergy spectrum of the hamiltonian with potential $V_0$ is given by an infinite equidistan ladder
$E_n=n+\frac{1}{2}$, where $n=0,1,...$, the corresponding eigenfunctions are 
\begin{equation}\label{eigen0}
 \psi_n(x) = \frac{1}{\pi^{1/4}\,\sqrt{2^n\,n!}}\,{\rm e}^{-x^2/2}h_n(x) \,,
\end{equation}
where $h_n(x)$ is the $n$-th Hermite polynomial.

On the other hand, the energy spectra of the supersymmetric partners of the harmonic oscillator $H_1$
consist in general of an isospectral subset $E_n=n+\frac{1}{2}$, $n=0,1,...$, 
whose corresponding eigenfunctions are
\begin{equation}
 \phi_n(x)=\frac{Q^+\psi_n(x)}{\sqrt{(E_n-\epsilon_1)...(E_n-\epsilon_k)}} \,,\quad n=0,1,...\,.                        
\end{equation}
However, in general, for a $k$-th order transformation, i.e. one where operator $Q^+$ is of $k$-th order,
$k$ new values $\epsilon_j$, $j=1,..,k$, may be added to the energy spectra of $H_1$,
associated to the eigenfunctions
\begin{equation}
 \phi_{\epsilon_j}\propto\frac{W(u_1,..,u_{j-1},u_{j+1},...,u_k)}{W(u_1,..,u_k)} \,,\quad j=1,...,k\,,
\end{equation}
respectively.
$W(u_1,..,u_k)$ is the wronskian of $k$ functions $u_j$, $j=1,..,k$, 
that solve the stationary Schr\"odinger equation $H_0u_j=\epsilon_ju_j$.
Although these $u_j$'s do not necessarily satisfy the boundary conditions 
of the eigenvalue problem for $H_1$.
Thus, the general form they take is given by
\begin{equation}
 u_j(x)={\rm e}^{-x^2/2}\left[
       \,_1F_1\left(\frac{1-2\epsilon_j}{4},\frac{1}{2},x^2\right)
         +2\nu x\,\frac{\Gamma(\frac{3-2\epsilon_j}{4})}
                   {\Gamma(\frac{1-2\epsilon_j}{4})}
                      \,_1F_1\left(\frac{3-2\epsilon_j}{4},\frac{3}{2},x^2\right)\right] \,, 
\end{equation}
and not (\ref{eigen0}).
$\Gamma(x)$ is the gamma function, 
$\,_1F_1(a,b,x)$ is the hypergeometric confluent function of $x$ with parameters $a$ and $b$,
and $\nu$ is a constant.

The potential in $H_1$ is 
\begin{equation}\label{V1}
 V_1(x) = \frac{x^2}{2} - \left[\ln W(u_1,...,u_k) \right]'' \,,
\end{equation}
where $'=\frac{{\rm d}}{{\rm d}x}$.
In order to have a non-singular transformation and, in turn, 
to not break the domain of definition of the eigenvalue problem, 
the wronskian $W(u_1,..,u_k)$ must not possess zeroes in the domain of definition of the problem,
which in this case is the whole real line $x\in\mathbb{R}$.
Let us note that the expression for $V_1$, eq. (\ref{V1}),
defines a multiparametric family of potentials.
The parameters are precisely the possibly added energy values $\epsilon_1,...,\epsilon_k$.

Supersymmetric transformations can be also used to realize the so-called \emph{factorization method},
that consists in identifying hamiltonians whose energy spectra can be obtained algebraically \cite{i41,ih51}.
First, consider the hermitian conjugate of equation (\ref{intertwining}),
\begin{equation}\label{hintertwining}
 QH_1 = H_0Q\,,
\end{equation}
where $Q=(Q^+)^\dagger$ is the hermitian conjugate of $Q^+$.
Then one can express the products of the intertwining operators $Q$ and $Q^+$ as the polynomials 
\begin{equation}\label{factorization}
QQ^+ = \prod_{i=1}^{k}(H_0-\epsilon_i)\,,\qquad 
Q^+Q = \prod_{i=1}^{k}(H_1-\epsilon_i)\,, 
\end{equation}
where $k$ is the differential order of $Q$ and $Q^+$, 
and $\epsilon_i$, $i=1,\dots,k$ are the roots of the operator polynomials in $H_0$ and $H_1$.
These are often called \emph{factorization energies}.

\section{Polynomial Heisenberg algebras}
Polynomial Heisenberg algebras (PHA) are deformations of the algebra of the harmonic oscillator \cite{cfnn04}.
These are defined by three operators, $H$, $L^+$ and $L^-$, together with the commutation relations
\begin{equation}\label{pha}
 [H,L^\pm] = \pm L^\pm \,, \qquad
 [L^-,L^+] = N(H+1)-N(H) = P_m(H) \,,
\end{equation}
where $N(H)=L^+L^-$ is the analogue of the number operator of the harmonic oscillator 
and $P_m(H)$ is a polynomial of degree $m$ of the hamiltonian $H$.

Notice that, indeed, the first commutation relations are the defining ones for ladder operators $L^\pm$, 
while the commutator between $L^+$ and $L^-$ characterize the deformation of the harmonic oscillator algebra.
The degree $m$ of the polynomial $P_m(H)$ defines the so-called degree of the PHA. 
If the differential order of the ladder operators is $m+1$ and the hamiltonian is of second differential order, 
as is usual in quantum mechanics, 
then the corresponding polynomial Heisenberg algebra is in general of degree $m$,
also denoted as $m$-PHA.
Indeed, 
\begin{equation}\label{0.4.3}
 N(H) = \prod_{i=1}^{m+1}(H-\mathcal{E}_i)\,,
\end{equation}
where $\mathcal{E}_i$ are the roots of the polynomial $N(H)$.

This algebras describe rather interesting physical systems,
since their energy spectra generally consists of $m+1$ equidistant sets of values (ladders).
To see how this happens consider the functions $\psi(x)$ in the kernel of the operator $L^-$, 
i.e., those that satisfy $L^-\psi=0$.
Then, $L^+L^-\psi=0$ and 
\begin{equation}\label{NH}
 \prod_{i=1}^{m+1}(H-\mathcal{E}_i)\psi=0.
\end{equation}

Now, from the relations (\ref{pha}) we obtain 
\begin{equation}\label{extremals}
 L^-H\psi=(H+1)L^-\psi=0,
\end{equation}
which in turn implies that the kernel of $L^-$ is invariant under the action of $H$.
Thus, by choosing the linearly independent functions $\psi_i$ generating the ker$(\,L^-)$ 
as the eigenfunctions of $H$ corresponding to the eigenvalues $\mathcal{E}_i$.
This are usually called \emph{extremal states} of $H$.
Finally, departing from each extremal state, 
the system described by the $m$-PHA possesses $m+1$ eigenenergy ladders, 
obtained from the repeated action of  $L^+$ on such states.

Examples of PHAs can be readily obtained 
by performing a supersymmetric transormation on the harmonic oscillator.
First, we note that the hamiltonian $H_1$, with potential given by (\ref{V1}),
has ladder operators defined in a natural way by 
\begin{equation}
 \ell^+=Q^+a^+Q\,,\quad \ell^-=Q^+a^-Q\,, 
\end{equation}
where $a^\pm=\left(x\mp \frac{{\rm d}}{{\rm d}x}\right)/\sqrt{2}$ 
are the usual ladder operators of the harmonic oscillator.
Then, while $Q^+$ and $Q$ are differential operators of $k$-th order, 
$\ell^+$ and $\ell^-$ are differential operators of $2k+1$-th order. 

The definition of $\ell^\pm$, together with ommutation relations between 
$H_0$ and $a^\pm$ for the harmonic oscillator, 
i.e. $[H_0,a^\pm]=\pm a^\pm$, imply 
\begin{equation}\label{0.3.7} 
 H_1 \ell^\pm = \ell^\pm (H_1\pm 1).
\end{equation}
Hence, $[\tilde{H},\ell^\pm]=\pm \ell^\pm$, 
and $\ell^\pm$ are indeed ladder operators for $H_1$.
However, the commutation relation between $\ell^+$ and $\ell^-$ satisfy 
\begin{equation}
 \ell^-\ell^+=\left(H-\frac{1}{2}\right)\prod_{i=1}^k(H-\epsilon_i)(H-\epsilon_i-1\,) \,.
\end{equation}
In contrast to the case of the harmonic oscillator,
where ladder operators satisfy $[a^-,a^+]=1$,
for its supersymmetric partners $[\ell^-,\ell^+]$ is a polynomial $P_m(\tilde H)$.

Based on the examples just mentioned, we can see that
supersymmetric transformations can be used to better understand PHAs.
The key idea that we will follow now is to employ the supersymmetric transformations 
to decompose the ladder operators into a product of first-order intertwining operators.
This will allow us to connect the PHAs to symmetric systems of equations 
characterized by extended affine Weyl groups.

\section{$A^{(1)}_n$ characterization of polynomial Heisenberg algebras}
Before tackling the general case, 
let us build some intuition about how the connection between PHAs and affine Weyl groups occur for low degrees.
This will also help us to obtain general quantum systems realizing the corresponding $m$-PHA.
As described in the previous section, PHAs are defined by the commutation relations (\ref{pha}), 
where $H=-\frac{1}{2}\frac{{\rm d}^2}{{\rm d}x^2}+V$ and $L^\pm$ are $m+1$-th order differential operators. 

\subsection{$0$-PHA}
The PHA with the lowest degree, i.e. zero, is obtained by considering first-order ladder operators
\begin{equation}\label{Lp0}
 L^+ = \frac{1}{2^{1/2}}\left(\frac{\rm d}{{\rm d}x}-f_1 \right) \,,
\end{equation}
\begin{equation}\label{Lm0}
 L^- = \frac{1}{2^{1/2}}\left(-\frac{\rm d}{{\rm d}x}-f_1 \right) \,,
\end{equation}
where $f_1=f_1(x)$ is a function of the coordinate $x$.
Thus, $f_1$ fixes $L^\pm$.

Now consider an auxiliar hamiltonian $H_2=H+\lambda$, where $\lambda\in\mathbb{R}$, intertwined with $H_1=H$ as
\begin{equation}\label{intert0}
 H_2L^+ = L^+H_1\,.
\end{equation}
Alternatively on could use the hermitian conjugate $H_1L^-=L^-H_2$.
This, by means of equation (\ref{factorization}), leads to the following factorizations:
\begin{equation}\label{factori0}
 H_1=L^-L^++\epsilon_1\,,\quad H_2=L^+L^-+\epsilon_1\,,
\end{equation}
recall that $\epsilon_1$ is a factorization energy. 
Using these factorizations in the intertwining (\ref{intert0}) 
yields the following expression for potential of $H$ in terms of $f_1$:
\begin{equation}\label{potential}
 V = f_1' + f_1^2 + \epsilon_1 \,,
\end{equation}
where we have used the notation $'=\frac{\rm d}{{\rm d}x}$.

We can see that, in general, 
$f_1$ specifies a realization of a $0$-PHA through equations (\ref{Lp0}), (\ref{Lm0}) and (\ref{potential}). 
However, equation (\ref{potential}) is not the only result from \label{intert0} and (\ref{factorization}).
The following equation for $f_1$ is also obtained:
\begin{equation}\label{f1-0}
 f_1' =  \lambda \,.
\end{equation}
This indeed shows that the general quantum system described by a $0$-PHA is given by a potential $V(x)$ quadratic in $x$.
The harmonic oscillator is the main representative of this family of physical systems.
A $0$-PHA is indeed the oscillator algebra where the commutation relation between $L^-$ and $L^+$ can be rescaled.

\subsection{$1$-PHA}
A first-degree PHA is obtained by considering second-order ladder operators
\begin{eqnarray}
 L^+ = Q_2^+Q_1^+ = \frac{1}{2}\left(\frac{{\rm d}}{{\rm d}x}-f_2 \right)\left(\frac{{\rm d}}{{\rm d}x}-f_1 \right) \,,\\
 L^- = Q_1^-Q_2^- = \frac{1}{2}\left(-\frac{{\rm d}}{{\rm d}x}-f_1 \right)\left(-\frac{{\rm d}}{{\rm d}x}-f_2 \right) \,,
\end{eqnarray}
where we have factorized $L^\pm$ in terms of first-order operators $Q_i^\pm$, $i=1,2$.
This decomposition is used to produce a series of intertwinings  
\begin{equation}\label{intert1}
 H_{3}Q_{2}^+ = Q_{2}^+H_{2}\,,\qquad H_{2}Q_{1}^+ = Q_{1}^+H_{1}\,,
\end{equation}
with auxiliar Hamiltonians $H_2$ and $H_3$, and $H_1 = H$.

We also consider the closure relation $H_{3}+\lambda=H_1$, where $\lambda\neq0$ is taken as a real number again.
Thus, we can wirte the folloign factorizations of the hamiltonians:
\begin{eqnarray}\label{factori1-1}
 H_1 &=& Q_1^-Q_1^++\epsilon_1\,, \\ \label{factori1-2}
 H_2 &=& Q_1^+Q_1^-+\epsilon_1 = Q_2^-Q_2^++\epsilon_2\,, \\ \label{factori1-3}
 H_3 &=& Q_2^+Q_2^-+\epsilon_2\,.
\end{eqnarray}
By using (\ref{factori1-1})-(\ref{factori1-3}) in (\ref{intert1}), we obtain, once more, 
that the potential in the hamiltonian $H$ is given by (\ref{potential}).
However, insead of equation (\ref{f1-0}) we obtain that functions $f_1$ and $f_2$ satisfy
\begin{eqnarray}\nonumber
 f_1' + f_2' &=&  f_1^2 - f_2^2 + 2(\epsilon_1-\epsilon_2) \,, \\\label{f1-1}
 f_1' + f_2' &=&  f_2^2 - f_1^2 + 2(\epsilon_2-\epsilon_1+\lambda) \,.
\end{eqnarray}

Furthermore, these lead to
\begin{eqnarray}
 (f_1+f_2)'=\lambda \,,\\
 f_1^2-f_2^2+2(\epsilon_1-\epsilon_2)=\lambda \,,
\end{eqnarray}
that, in trun, yield an expression for $f_2$ in terms of $f_1$:
\begin{equation}
 f_2 = \lambda x + c_0 - f_1 \,,
\end{equation}
where $c_0$ is an integration constant;
and a general expression for $f_1$:
\begin{equation}
 f_1 = \lambda x + c_0  + \frac{ 2(\epsilon_2-\epsilon_1) + \lambda}{\lambda x + c_0}\,.
\end{equation}

In a similar fashion as for a $0$-PHA, 
$f_1$ specifies a realization of a $1$-PHA.
Then, by explicitly writing the resulting  expression for the potential of $H$, 
we can see that the general quantum system described by a $1$-PHA possesses potentials of the form
\begin{equation}
 V(x) =  \left(\frac{2\epsilon_2-2\epsilon_1+\lambda }{c_0+\lambda  x}+c_0+\lambda x\right)^2
         -\frac{\lambda (2\epsilon_2-2\epsilon_1+\lambda )}{(c_0+\lambda x)^2}+\epsilon_1+\lambda \,,
\end{equation}
with the radial oscillator as the main representative of such systems \cite{cfnn04}.

Evenmore, system (\ref{f1-1}) admits rational transformations characterized 
by the extented affine Weyl group of type $A_1^{(1)}$; 
namely, $\tilde{W}(A_1^{(1)})=\langle s_0, s_1,\pi\rangle$.
These transformations are given by
\begin{eqnarray}
 s_j(f_j)     = f_j     + \frac{\alpha_j}{f_i+f_{j+1}} \,,&& s_j(\alpha_j)       = -\alpha_j \,, \\
 s_j(f_{j+1}) = f_{j+1} - \frac{\alpha_j}{f_i+f_{j+1}} \,,&& s_j(\alpha_{j\pm1}) = \alpha_{j\pm1}+2\alpha_j \,, \\
 \pi(f_j)     = f_{j+1} \,,                               && \pi(\alpha_j)       = \alpha_{j+1} \,,
\end{eqnarray}
where $i,j = 0, 1$ and $\alpha_i = 2(\epsilon_{i+1}-\epsilon_{i+2})$.
Notice that the closure relation previously imposed implies that $\epsilon_3 + \lambda = \epsilon_1$.

Recall that $\tilde{W}(A_{n-1}^{(1)})$ is the group defined by the generators 
$s_0$, $s_1$, ..., $s_{n-1}$ and $\pi$, together with the fundamental relations 
\begin{eqnarray}
 s_i^2 &=& 1\,, \\
 (s_is_j)^2 &=& 1 \quad (j\neq i\,,i\pm 1)\,, \\
 (s_is_j)^3 &=& 1 \quad (j=i\pm 1)\,, \\
 \pi s_i &=& s_{i+1}\pi \quad (i=0,1,...,n-1)\,, \\
 \pi^n &=& 1 \,.
\end{eqnarray}

\subsection{$2$-PHA}
Now let us consider the case of third order ladder operators
\begin{eqnarray}
 L^+ = A_3^+A_2^+A_1^+ = \frac{1}{2^{3/2}}\left(\frac{{\rm d}}{{\rm d}x}-f_3 \right)\left(\frac{{\rm d}}{{\rm d}x}-f_2 \right)\left(\frac{{\rm d}}{{\rm d}x}-f_1 \right) \,, \\
 L^- = A_1^-A_2^-A_3^- = \frac{1}{2^{3/2}}\left(-\frac{{\rm d}}{{\rm d}x}-f_1 \right)\left(\frac{-{\rm d}}{{\rm d}x}-f_2 \right)\left(-\frac{{\rm d}}{{\rm d}x}-f_3 \right) \,,
\end{eqnarray}
in a second degree PHA.
Again, we have factorized both ladder operators in terms of first order operators $Q_i^\pm$, $i=1,2,3$,
and use them to produce the intertwining relations 
\begin{equation}\label{intert2}
 H_{4}Q_{3}^+ = Q_{3}^+H_{3}\,,\quad H_{3}Q_{2}^+ = Q_{2}^+H_{2}\,,\quad H_{2}Q_{1}^+ = Q_{1}^+H_{1}\,,
\end{equation}
where $H_2$, $H_3$ and $H_4$ are auxiliar Hamiltonians, and $H_1 = H$.

This time the closure ralation is set to be $H_4+\lambda=H_1$, $\lambda\in\mathbb{R}$.
The resulting factorizations of the hamiltonians are given as 
\begin{eqnarray}\nonumber
 H_1=Q_1^-Q_1^++\epsilon_1\,, \quad H_2=Q_1^+Q_1^-+\epsilon_1=Q_2^-Q_2^++\epsilon_2\,, \\\label{factori2}
 H_4=Q_3^+Q_3^-+\epsilon_3\,, \quad H_3=Q_2^+Q_2^-+\epsilon_2=Q_3^-Q_3^++\epsilon_3\,.
\end{eqnarray}

Equations (\ref{factori2}) and (\ref{intert2}) can be combined to obtain equation (\ref{potential}),
as well as the following system of equations: 
\begin{eqnarray}\nonumber
 f_1' + f_2' &=&  f_1^2 - f_2^2 + 2(\epsilon_1-\epsilon_2)\,, \\\label{f1-2}
 f_2' + f_3' &=&  f_2^2 - f_3^2 + 2(\epsilon_2-\epsilon_3)\,, \\\nonumber
 f_3' + f_1' &=&  f_3^2 - f_1^2 + 2(\epsilon_3-\epsilon_1+\lambda)\,.
\end{eqnarray}
Adding these, we can obtain the equation 
\begin{equation}
 (f_1+f_2+f_3)' = \lambda \,.
\end{equation}
However, this time, obtaining the equations that functions $f_i$, $i=1,2,3$, 
satisfy requires some lenghty calculations that can be found in \cite{cfnn04,bf14}.
The result yields 
\begin{eqnarray}
 2f_2 = (\lambda x +c_0 -f_1) - \frac{f_1' +2(\epsilon_2-\epsilon_3) -\lambda}{\lambda x +c_0 -f_1} \,,\\
 2f_3 = (\lambda x +c_0 -f_1) + \frac{f_1' +2(\epsilon_2-\epsilon_3) -\lambda}{\lambda x +c_0 -f_1} \,,
\end{eqnarray}
that show how to obtain $f_2$ and $f_3$ from $f_1$,
while $f_1$ satisfies is a second order non-linear differential equation that,
upon the sustitution $f_1=g(x)-\lambda x+c_0$, 
can be shown to be equivalent to the Painlev\'e IV equation
\begin{equation}\label{Piv}
 g'' = \frac{(g')^2}{2g}+\frac{3}{2}g^3+4xg^2+2(x^2-b_0)g+\frac{b_1}{g} \,,
\end{equation}
where $b_0$ and $b_1$ are constants.

One can see that, similarly to the previous cases, 
$V$, $f_2$ and $f_3$ can be obtained from $f_1$, 
that in trun can be obtained by solving the Painlev\'e IV equation.
On the other hand, this time the general quantum system described by a $2$-PHA 
possesses a potential of the form
\begin{equation}
 V(x) = \frac{f_1^2}{2} -\frac{f_1'}{2} +xf_1 +\frac{x^2}{2} +c \,,
\end{equation}
where $c$ is a constant.

The connection between system (\ref{f1-2}) and Painlev\'e IV equation resides in the fact that 
the former is precisely the symmetric form of the latter. 
System (\ref{f1-2}) possesses a set of B\"acklund transformations given by 
\begin{eqnarray}\label{backlund1}
 s_j(f_j)     = f_j     + \frac{\alpha_j}{f_i+f_{j+1}} \,,&& s_j(\alpha_j)       = -\alpha_j \,, \\\label{backlund2}
 s_j(f_{j+1}) = f_{j+1} - \frac{\alpha_j}{f_i+f_{j+1}} \,,&& s_j(\alpha_{j\pm1}) = \alpha_{j\pm1}+\alpha_j \,, \\\label{backlund3}
 \pi(f_j)     = f_{j+1} \,,                               && \pi(\alpha_j)       = \alpha_{j+1} \,,
\end{eqnarray}
where $i,j = 0,1,2$ and $\alpha_i = 2(\epsilon_{i+1}-\epsilon_{i+2})$,
with $\epsilon_4 + \lambda = \epsilon_1$.
Notice the change of a coefficient in the transformation $s_j(\alpha_{j\pm1})$ with respect to the previous case.
These transformations are characterized by the extented affine Weyl group 
$\tilde{W}(A_2^{(1)})=\langle s_0, s_1,s_2,\pi\rangle$.

\subsection{$3$-PHA}
Now that the pattern has been emerging as we reviewed the cases of PHAs of degree $0$, $1$ and $2$,
let us just state the results for a third degree PHA.  
First we consider fourth order ladder operators
\begin{eqnarray}
 L^+ = Q_4^+Q_3^+Q_2^+Q_1^+ \,, \quad L^- = Q_1^-Q_2^-Q_3^-Q_4^- \,, 
\end{eqnarray}
where $Q_i^\pm=\frac{1}{\sqrt{2}}\left(\pm\frac{{\rm d}}{{\rm d}x}-f_i\right)$, $i=1,2,3,4$
are used to produce intertwinings 
\begin{equation}\label{intert2}
 H_{5}Q_{4}^+ = Q_{4}^+H_{4}\,,\quad H_{4}Q_{3}^+ = Q_{3}^+H_{3}\,,\quad H_{3}Q_{2}^+ = Q_{2}^+H_{2}\,,\quad H_{2}Q_{1}^+ = Q_{1}^+H_{1}\,,
\end{equation}
among the auxiliar hamiltonians $H_i$, $i=1,2,3,4$, such that $H_1=H$ and $H_5+\lambda=H_1$, $\lambda\neq0$.
By means of the factorization method equation (\ref{potential}) holds and we obtain the set of equations  
\begin{eqnarray}\nonumber
 f_1' + f_2' &=&  f_1^2 - f_2^2 + 2(\epsilon_1-\epsilon_2)\,, \\\nonumber
 f_2' + f_3' &=&  f_2^2 - f_3^2 + 2(\epsilon_2-\epsilon_3)\,, \\\label{f1-3}
 f_3' + f_4' &=&  f_3^2 - f_4^2 + 2(\epsilon_3-\epsilon_4)\,, \\\nonumber
 f_4' + f_1' &=&  f_4^2 - f_1^2 + 2(\epsilon_4-\epsilon_1+\lambda)\,.
\end{eqnarray}
This set of equations can be reduced to the Painlev\'e V equation \cite{cfnn04,bf14}
{\footnotesize
\begin{equation}\label{PV}
 \frac{{\rm d}^2w}{{\rm d}z^2}= \left(\frac{1}{2w}+\frac{1}{w-1}\right)\left(\frac{{\rm d}w}{{\rm d}z}\right)^2
                  -\frac{1}{z}\frac{{\rm d}w}{{\rm d}z} +\frac{(w-1)^2}{z^2}\left(c_1w+\frac{c_2}{w}\right) +c_3\frac{w}{z}
                   +c_4\frac{w(w+1)}{w-1} \,,
\end{equation}}
where $c_i$, $i=1,...,4$ are constants and a change of variables, $f_1=f_1(w)$ and $x=x(z)$, is used. 

Quantum systems described by a $3$-PHA can be obtained by means of equation (\ref{potential}),
together with particular solutions of the fifth Painlev\'e equation and $f_1=f_1(w)$, $x=x(z)$.
Indeed, system (\ref{f1-3}) is the symmetric form of Painlev\'e V equation,
and its rational B\"acklund transformations are characterized by the extented affine Weyl group 
$\tilde{W}(A_3^{(1)})=\langle s_0, s_1,s_2,s_3,\pi\rangle$.
On the other hand, 
their rational B\"acklund transformations, eqs. (\ref{backlund1})-(\ref{backlund3}),
where, in the general case, $i,j = 0,...,3$ and $\alpha_i = 2(\epsilon_{i+1}-\epsilon_{i+2})$.
Of course, $\epsilon_5 + \lambda = \epsilon_1$.

\subsection{Higher order PHA}
In this subsection we present the general case of an $m$-PHA, $m>1$.
We will build on the cases of low degree PHAs presented in the previous sections.
Again, consider the defining relations of an $m$-PHA (\ref{pha}), 
such that $H=-\frac{1}{2}\frac{{\rm d}^2}{{\rm d}x^2}+V$ and
\begin{eqnarray}
 L_{m+1}^+ &=& \prod_{i=1}^{m+1}Q_{m+2-i}^+ \,, \\
 L_{m+1}^- &=& \prod_{i=1}^{m+1}Q_i^- \,,
\end{eqnarray}
where $Q_i^\pm=\frac{1}{2^{1/2}}\left(\pm\frac{{\rm d}}{{\rm d}x}-f_i\right)$, $f_i\in\mathbb{R}$, $i=1,...,m+1$. 
Operators $Q_i^\pm$ fulfill the intertwining relations
\begin{equation}
 H_{j}Q_{j}^- = Q_{j}^-H_{j+1}\,,\qquad H_{j+1}Q_{j}^+ = Q_{j}^+H_{j}\,,
\end{equation}
where $H_j$ are auxiliar Hamiltonians and $H_1=H$.
This, in turn, leads to the factorizations
\begin{equation}
 H_j=Q_j^-Q_j^++\epsilon_j\,,\qquad H_{j+1}=Q_j^+Q_j^-+\epsilon_j\,,
\end{equation}
These, together with the closure relation $H_{m+2}+\lambda=H_1$, lead to  
\begin{equation}
 V(x) = f_1' + f_1^2 + \epsilon_1 \,.
\end{equation}
and the system of equations
\begin{eqnarray}\nonumber
 f_1' + f_2' &=&  f_1^2 - f_2^2 + 2(\epsilon_1-\epsilon_2) \,,\\ \label{sse}
             &\vdots& \\ \nonumber
 f_m' + f_{m+1}' &=&  f_m^2 - f_{m+1}^2 + 2(\epsilon_m-\epsilon_{m+1}) \,, \\ \nonumber
 f_{m+1}' + f_1' &=&  f_{m+1}^2 - f_1^2 + 2(\epsilon_{m+1}-\epsilon_1+\lambda) \,.
\end{eqnarray}
One immediate result from this system is the equation
\begin{eqnarray}
 (f_1+f_2+\hdots+f_m+f_{m+1})'=\lambda \,.
\end{eqnarray}

Rational B\"acklund transformations of system (\ref{sse}) are characterized 
by the extented affine Weyl group of type $A_m^{(1)}$, 
i.e., $\tilde{W}(A_m^{(1)})=\langle s_0, s_1, \cdots s_m,\pi\rangle$.
These B\"acklund transformations take the explicit form of (\ref{backlund1})-(\ref{backlund3}),
that we reproduce here for completeness:
\begin{eqnarray}\label{backlund1}
 s_j(f_j)     = f_j     + \frac{\alpha_j}{f_i+f_{j+1}} \,,&& s_j(\alpha_j)       = -\alpha_j \,, \\\label{backlund2}
 s_j(f_{j+1}) = f_{j+1} - \frac{\alpha_j}{f_i+f_{j+1}} \,,&& s_j(\alpha_{j\pm1}) = \alpha_{j\pm1}+\alpha_j \,, \\\label{backlund3}
 \pi(f_j)     = f_{j+1} \,,                               && \pi(\alpha_j)       = \alpha_{j+1} \,,
\end{eqnarray}
where $i,j = 0,1,...,m-1$ and $\alpha_i = 2(\epsilon_{i+1}-\epsilon_{i+2})$.
The closure relation implies that $\epsilon_{m+2} + \lambda = \epsilon_1$.

Results thus far show the explicit way in which 
extended affine Weyl groups characterize polynomial Heisenberg algebras.
Thus, one could talk about the \emph{B\"acklund transformations} of an $m$-PHA
characterized by the extended affine Weyl group of type $A_m^{(1)}$.
Concrete realizations of a PHA of a fixed degree can be obtained 
by performing said transformations.

The fact that for $m=0,1$ PHAs do not follow the characterization 
by extended affine Weyl groups seems to derive from the facts that 
1) the cyclic dressing chains (\ref{sse}) possess different behaviors depending on the parity of $m$, and 
2) for each of these clases $m=0,1$ represent the trivial cases, respectively.
Thus, for $m>3$, the system (\ref{sse}) is regarded as higher order generalizations of 
either the fourth or fifth Painlev\'e equations, according to $m$'s parity.

As mentioned in the introduction, 
the connection between PHAs and non-linear differential equations can be used to study solutions of the latter, 
here presented in symmetric form.
Now, by means of the rational transformations,
work has been done trying to connect rational solutions of such equations;
in particular, for Painlv\'e IV and V.
On the other hand, nonrational hierarchies of solutions have also been obtained, e.g,
confluent hypergemoetric function hierachy and error function hierachy \cite{bf14}.
Results presented here can be used to study connections among solutions in these hierarchies,
by using nonrational instead of rational seed solutions, for example.

\section{Final remarks}
We have seen that polynomial Heisenberg algebras of $m$-th order ($m$-PHA)
are deformations of the oscillator algebra,
where commutation relations between the hamiltonian and the ladder operators 
are the usual ones as in the harmonic oscillator.
However, the commutation relation between the ladder operators 
is an $m$-th degree polynomial of of the hamiltonian.
By means of appropriate supersymmetric transformations, 
one can obtain a symmetric system of equations (\ref{sse}) equivalent to the $m$-PHA known as cyclic dressing chains. 
Upon solving it, we can obtain the most general quantum system described by the polynomial Heisenberg algebra.

Even more, the rational B\"acklund transformations of system (\ref{sse}) are characterized 
by the extented affine Weyl group of type $A_m^{(1)}$, 
i.e., $\tilde{W}(A_m^{(1)})=\langle s_0, s_1, \cdots s_m,\pi\rangle$.
For $m=1$, system (\ref{sse}) admits rational B\"acklund transformations characterized by the extented affine Weyl group of type $A_1^{(1)}$.
For $m=2$, (\ref{sse}) is known as the symmetric form of Painlev\'e IV equation; 
whose rational B\"acklund transformations and characterized by the extented affine Weyl group of type $A_2^{(1)}$.
For $m=3$, system (\ref{sse}) is known as the symmetric form of Painlev\'e V equation;
and its rational B\"acklund transformations are characterized by the extented affine Weyl group of type $A_3^{(1)}$.

We can see that as the degree of the polynomial Heisenberg algebra increases,
solving the system (\ref{sse}) becomes more difficult, e.g, for $m>3$,
the non-linear differential equation equivalent to (\ref{sse}) is of order greater than two.
Connections described here may help to better study quantum systems described by PHAs, on one hand, 
and (non-linear) differential equations and affine Weyl groups, on the other.

\section{Acknowledgments}
The author would like to greatly thank and show appreciation to Dr. Yang Shi 
for pointing towards the outstanding work of M. Noumi on symmetries in the study of Painlev\'e equations.



\end{document}